\documentclass{aip-cp}

\usepackage[numbers]{natbib}
\usepackage{rotating}
\usepackage{graphicx}

\newcommand{\be}{\begin{equation}}
\newcommand{\ee}{\end{equation}}
\newcommand{\ba}{\begin{eqnarray}}
\newcommand{\ea}{\end{eqnarray}}
\newcommand{\apj}{Astrophys. J.}
\newcommand{\apjl}{Astrophys. J. Lett.}
\newcommand{\aap}{Astron. \& Astrophys.}
\newcommand{\apss}{Astrophys. \& Sp. Sci.}
\newcommand{\iaucirc}{IAU Circ.}
\newcommand{\mnras}{Mon. Not. Roy. Astron. Soc.}
\newcommand{\nat}{Nature}
\newcommand{\procspie}{Proc. SPIE Conf. Ser.}
\newcommand{\physrep}{Phys. Rep.}
\newcommand{\prc}{Phys. Rev. C}
\newcommand{\ssr}{Sp. Sci. Rev.}
\newcommand{\Chandra}{\textit{Chandra}}
\newcommand{\Ts}{\ensuremath{T_{\rm s}}}
\newcommand{\Tb}{\ensuremath{T_{\rm b}}}
\newcommand{\tiso}{\ensuremath{t_{\rm relax}}}
\newcommand{\Msun}{\ensuremath{M_{\rm Sun}}}
\newcommand{\NH}{\ensuremath{N_{\rm H}}}

\begin{document}

\title{ Cooling of the Cassiopeia A Neutron Star and the \\ Effect of Diffusive Nuclear Burning}

\author[aff1,aff2,aff3]{Wynn C.G. Ho\corref{cor1}}
\author[aff2]{M.J.P. Wijngaarden}
\author[aff4]{Philip Chang}
\author[aff5]{Craig O. Heinke}
\author[aff6]{Dany Page}
\author[aff6]{Mikhail Beznogov}
\author[aff7]{Daniel J. Patnaude}

\affil[aff1]{Department of Physics and Astronomy, Haverford College, 370 Lancaster Avenue, Haverford, PA 19041, USA}
\affil[aff2]{Mathematical Sciences and STAG Research Centre, University of Southampton, SO17 1BJ Southampton, UK}
\affil[aff3]{Physics and Astronomy, University of Southampton, SO17 1BJ Southampton, UK}
\affil[aff4]{Department of Physics, University of Wisconsin-Milwaukee, 1900 E. Kenwood Blvd., Milwaukee, WI 53211, USA}
\affil[aff5]{Department of Physics, University of Alberta, CCIS 4-181, T6G 2E1 Edmonton, Alberta, Canada}
\affil[aff6]{Instituto de Astronom\'ia, Universidad Nacional Aut\'onoma de M\'exico, Mexico City, CDMX 04510, Mexico}
\affil[aff7]{Smithsonian Astrophysical Observatory, Cambridge, MA 02138, USA}
\corresp[cor1]{Corresponding author: wynnho@slac.stanford.edu}

\maketitle

\begin{abstract}
The study of how neutron stars cool over time can provide invaluable
insights into fundamental physics such as the nuclear equation of state
and superconductivity and superfluidity.
A critical relation in neutron star cooling is the one between observed
surface temperature and interior temperature. This relation is determined
by the composition of the neutron star envelope and can be influenced by
the process of diffusive nuclear burning (DNB).
We calculate models of envelopes that
include DNB and find that DNB can lead to a rapidly changing envelope
composition which can be relevant for understanding the long-term cooling
behavior of neutron stars.  We also report on analysis of the latest
temperature measurements of the young neutron star in the Cassiopeia~A
supernova remnant.  The 13 \Chandra\ observations over 18 years show that
the neutron star's temperature is decreasing at a rate of 2--3\% per decade,
and this rapid cooling can be explained by the presence of a proton
superconductor and neutron superfluid in the core of the star.
\end{abstract}

\section{INTRODUCTION TO NEUTRON STAR COOLING THEORY}

Neutron stars are born in the supernova explosion of a massive star
and begin their lives very hot
(with thermal energies $kT\gg1\mbox{ MeV}$ or
temperatures $T$ greatly exceeding $10^{10}\mbox{ K}$).
They cool rapidly due to neutrino emission and later on by photon
emission \cite{lattimeretal94,tsuruta98,yakovlevetal99,potekhinetal15}.
Neutrino emission processes depend on physics at densities that exceed
nuclear saturation
(at baryon density $n_{\rm nuc}\approx0.16\mbox{ fm$^{-3}$}$ or
mass densities $\rho_{\rm nuc}\approx2.8\times10^{14}\mbox{ g cm$^{-3}$}$).
The state of matter is uncertain at these densities, e.g.,
exotica such as a neutron superfluid and proton superconductor and
hyperons, axions, and deconfined quarks could be present
\cite{lattimer12,oerteletal17}.
By comparing how neutron stars cool theoretically over time with
observations of neutron stars of different ages, one can extract
and understand properties of fundamental physics.

Let us consider the basic two equations that govern neutron star cooling.
Here for simplicity, we ignore relativistic terms, extra sources of heating,
magnetic fields, etc. (see \cite{potekhinetal15}, for more details).
Then the equations of energy balance and heat flux are
\ba
\frac{1}{4\pi r^2}\frac{\partial L_r}{\partial r} &=&
 -C\frac{\partial T}{\partial t}-\varepsilon_\nu \label{eq:energy} \\
\frac{L_r}{4\pi r^2} &=& -K\frac{\partial T}{\partial r},
\ea
where $L_r$ is luminosity at radius $r$, $C$ is heat capacity,
$\varepsilon_\nu$ is neutrino emissivity, and $K$ is thermal conductivity.
The microphysics inputs that determine how a neutron star cools are $C$,
$\epsilon_\nu$, and $K$ (see \cite{yakovlevetal99,gnedinetal01,hoetal12}).
After a thermal relaxation time of
\be
\tiso \sim (C/K)R_{\rm crust}^2 \sim 10-100\mbox{ yr},
\ee
where $R_{\rm crust}$ is crust thickness,
the temperature gradient becomes small, such that the interior is
effectively isothermal, and thermal evolution is determined solely by
Equation~(\ref{eq:energy}), which can be further simplified by an
integration over volume.
A final important ingredient of neutron star cooling is the outer layers,
or envelope, of the neutron star crust, which prescribes the outer boundary
condition for the solution of the cooling equations.
Properties of the envelope, such as elemental composition, determine
the transition between temperature at the bottom of the envelope
(at $\sim 10^8-10^{10}\mbox{ g cm$^{-3}$}$) and temperature at the surface,
(i.e. $\Tb$--$\Ts$),
where surface temperature corresponds to that measured using telescopes
such as \Chandra\ and \textit{XMM-Newton}.

\begin{figure}[htb]
\centerline{\includegraphics[width=0.5\textwidth]{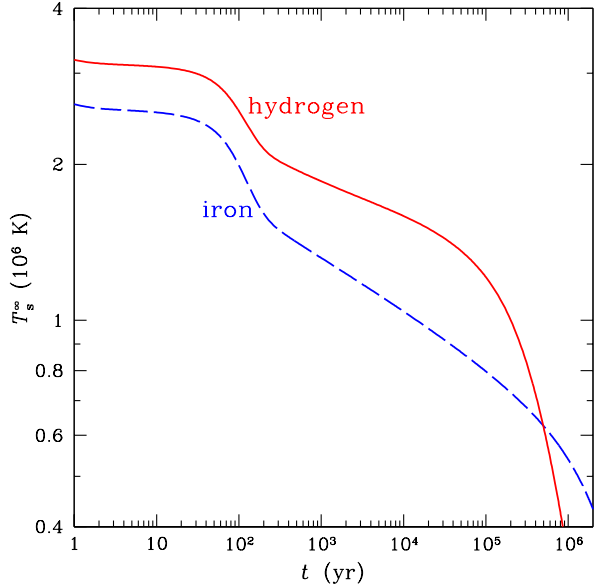}}
\caption{
Neutron star (redshifted) surface temperature $\Ts^\infty$ as a function
of age.  Solid line is for a cooling simulation which uses a hydrogen
envelope, while dashed line is one that uses an iron envelope.
\label{fig:cool}
}
\end{figure}

Figure~\ref{fig:cool} illustrates some of the basic features of a cooling
neutron star.
For instance, there are three stages of neutron star cooling.
At early times ($t<\tiso$), there is a period of thermal relaxation
when $\Ts\sim\mbox{constant}$.  The neutron star core is cooling rapidly
because neutrino emission is more effective at high densities.
Finite thermal conductivity of lower density regions,
such as the crust, delays when this temperature drop is seen at the
surface.
After this cooling wave from the core reaches the surface at $t\sim\tiso$,
we see a sharp decline of $\Ts$.
For standard neutrino emission processes
(such as the modified Urca reaction, $n+n\rightarrow n+p+e+\bar{\nu}_{\rm e}$,
which scales as $\varepsilon_\nu\propto T^8$), the interior temperature
is approximately isothermal and decreases as $\Tb\propto t^{-1/6}$.
Once temperature decreases to the point when neutrino emission becomes
less efficient than photon emission (at $t>10^5\mbox{ yr}$), the
latter becomes the dominant cooling process, and temperature decreases
as $\Tb\propto t^{-1/2}$.

We note here that,
when the interior temperature drops below (density-dependent) critical
temperatures for proton superconductivity and neutron ($^1S_0$ or $^3P_2$)
superfluidity, the accompanying phase transition causes two competing
effects on cooling.
First, superconducting protons and superfluid neutrons no longer participate
in standard neutrino emission processes such as the modified Urca reactions,
and thus cooling is suppressed (there is also a reduction of heat capacity;
\cite{yakovlevetal99}).
Second, formation and breaking of superfluid or superconducting Cooper
pairs lead to a new source of neutrino emission
(e.g., $n+n\rightarrow [nn]+\nu+\bar{\nu}$), which enhances cooling
\cite{yakovlevetal99}.

\section{ENVELOPE/ATMOSPHERE EVOLUTION AND DIFFUSIVE NUCLEAR BURNING}

Figure~\ref{fig:cool} also shows the dependence of
observed surface temperature $\Ts$ on envelope composition.
Since thermal conductivity is higher for light elements ($K\propto 1/Z$),
a light element envelope is more transparent to the hot core than a
heavy element envelope.
An outer layer composed of hydrogen could exist if accretion occurs after
neutron star formation.
Alternatively, a helium or carbon composition could result from diffusive
nuclear burning on the neutron star surface (see next).
Finally, a heavy element outer layer would be present if no accretion
takes place or if all lighter elements are consumed by nuclear reactions.
Each of these compositions gives a different $\Ts$--$\Tb$ relation,
e.g., $\Ts/10^6\mbox{ K}\sim 2(\Tb/10^8\mbox{ K})^{17/28}$ for hydrogen
\cite{potekhinetal97,potekhinetal03} and
$\Ts/10^6\mbox{ K}\sim (\Tb/10^8\mbox{ K})^{11/20}$ for iron
\cite{gudmundssonetal83}.  However, these works only consider
envelope compositions that do not change with time.
Recently, \cite{beznogovetal16} develop $\Ts$--$\Tb$ relations for
mixed compositions, including H--He, He--C, and C--Fe, and find that
diffusion is slow enough to not affect these relations.
\cite{beznogovetal16b} then use these relations to conduct cooling
simulations.
However these works do not consider the effects of diffusive nuclear
burning (DNB).

\begin{figure}[htb]
\centerline{\includegraphics[width=0.5\textwidth]{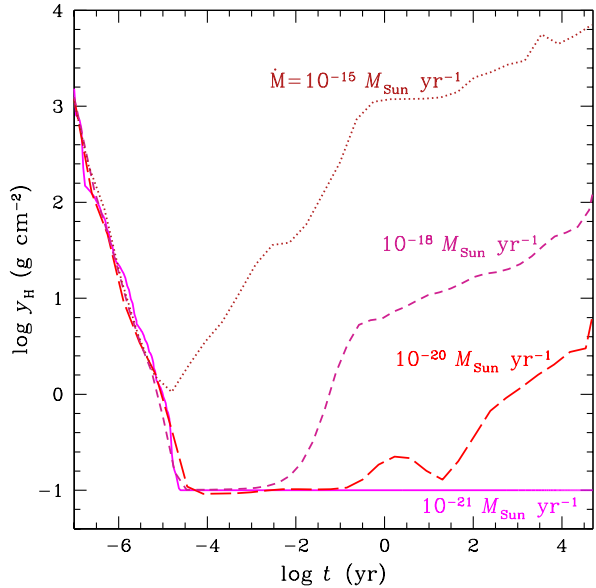}}
\caption{
Hydrogen column density $y_{\rm H}$ as a function of time.
An initial column of $y_{\rm H}=10^6\mbox{ g cm$^{-2}$}$
and a minimum column of $y_{\rm H}=0.1\mbox{ g cm$^{-2}$}$ are assumed.
Curves are for different mass accretion rates $\dot{M}$.
\label{fig:dnb}
}
\end{figure}

DNB occurs when temperatures are high enough for nuclear burning to
take place in a diffusive tail that extends from the surface into much
higher density layers \cite{chiusalpeter64,rosen68,rosen69,changbildsten03}.
This light element current can easily deplete surface layers of
hydrogen in $<10^5\mbox{ yr}$ \cite{changbildsten03,changbildsten04}
and helium in $<10^3\mbox{ yr}$ \cite{changetal10}.
In \cite{wijngaardenetal19}, we account for DNB in H-He and He-C envelopes
in calculating new $\Ts$--$\Tb$ relations (for H-C, see \cite{wijngaardenetal}),
which we use in neutron star cooling simulations.
Figure~\ref{fig:dnb} shows examples of how the amount of hydrogen at
the surface evolves for a H-C envelope, where we allow for an isolated
neutron star to accrete from its surrounding environment
(e.g., if the environment has a number density of $1\mbox{ cm$^{-3}$}$
and the neutron star moves at a velocity of $20\mbox{ km s$^{-1}$}$,
then $\dot{M}\sim 10^{-15}\mbox{ \Msun\ yr$^{-1}$}$;
see \cite{wijngaardenetal19}, for details).
We see that initial hydrogen is consumed within an hour due to
high temperatures present soon after neutron star formation.
For accretion rates $\dot{M}>10^{-20}\mbox{ \Msun\ yr$^{-1}$}$,
the amount of hydrogen builds up again, and
there is a sufficient amount to produce at least a hydrogen atmosphere
after a few hundred years.
Such a neutron star would be observed to have a hydrogen spectrum,
while a neutron star accreting at a lower $\dot{M}$ would have a carbon
spectrum \cite{potekhin14}.
Whether we could detect a near real-time change in atmosphere composition
is unknown.

\section{COOLING OF THE CASSIOPEIA A NEUTRON STAR}

\begin{figure}[!tb]
\centerline{\includegraphics[width=0.5\textwidth]{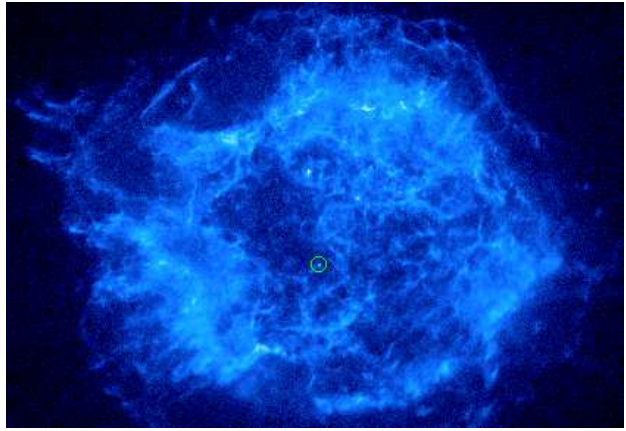}}
\caption{
\Chandra\ image (ObsID 14481) of the Cassiopeia~A supernova remnant.
Circle near the remnant center indicates the neutron star position.
\label{fig:casa}
}
\end{figure}

\begin{figure}[!tb]
\centerline{\includegraphics[width=0.5\textwidth]{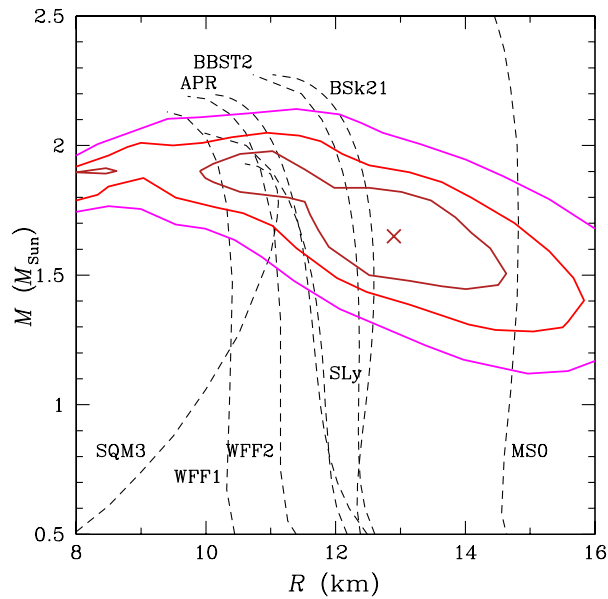}}
\caption{
Neutron star mass $M$ and radius $R$.
Solid lines are 1$\sigma$, 90\%, and 99\% confidence contours of
$M$--$R$ obtained from fitting \Chandra\ spectra of the Cassiopeia~A
neutron star.
The cross marks the best-fit $M=1.65\,\Msun$ and $R=12.9\mbox{ km}$.
Dashed lines indicate $M$--$R$ relation obtained from various
theoretical nuclear EOSs, in particular
APR \cite{akmaletal98}, BBST2 \cite{baldoetal14}, BSk21 \cite{potekhinetal13},
MS0, SQM3, WFF1, WFF2 (see \cite{lattimerprakash01}, and references therein),
and SLy \cite{douchinhaensel01}.
\label{fig:mr}
}
\end{figure}

The neutron star in the Cassiopeia~A supernova remnant is one of the
most important neutron stars because it is the youngest known, at an
age of $\approx$340~yr.
This estimated age comes from a supernova date of $1681\pm19$, which
is derived from studying the expansion of the supernova remnant
\cite{fesenetal06}.
While the remnant has been known about for a long time, the compact
object associated with the remnant and near its center
(see Figure~\ref{fig:casa}) was only discovered with the launch of
\Chandra\ and use of its unparalleled spatial resolution
\cite{tananbaum99}.
Although no pulsations are detected from the compact object
\cite{murrayetal02,halperngotthelf10}, its identification as a neutron
star with a carbon atmosphere can be made from its spectrum \cite{hoheinke09}.
Figure~\ref{fig:mr} shows neutron star mass $M$ and radius $R$ confidence
contours obtained from fitting the star's spectra
(see \cite{wijngaardenetal19}, for details)
and assuming a distance of 3.4~kpc \cite{reedetal95}.
It is important to note that, while the best-fit values are $M=1.65\Msun$
and $R=12.9\mbox{ km}$, these values are not strongly constrained,
and the uncertainty region is broad and encompasses many nuclear EOSs.
A notable effect of this $M$--$R$ uncertainty is on temperatures derived
from spectra (see next), which include fitting for gravitational redshift
($\propto 2GM/Rc^2$).
As a result, normalization of the absolute value of temperature
depends on mass and radius. 
However relative temperatures do not, and neither does the observed
cooling rate.
On the other hand, theoretical modeling of the observed cooling rate
depends on mass and radius
(see, e.g., \cite{pageetal11,shterninetal11,hoetal15}).

\begin{figure}[!tb]
\centerline{
\includegraphics[width=0.49\textwidth]{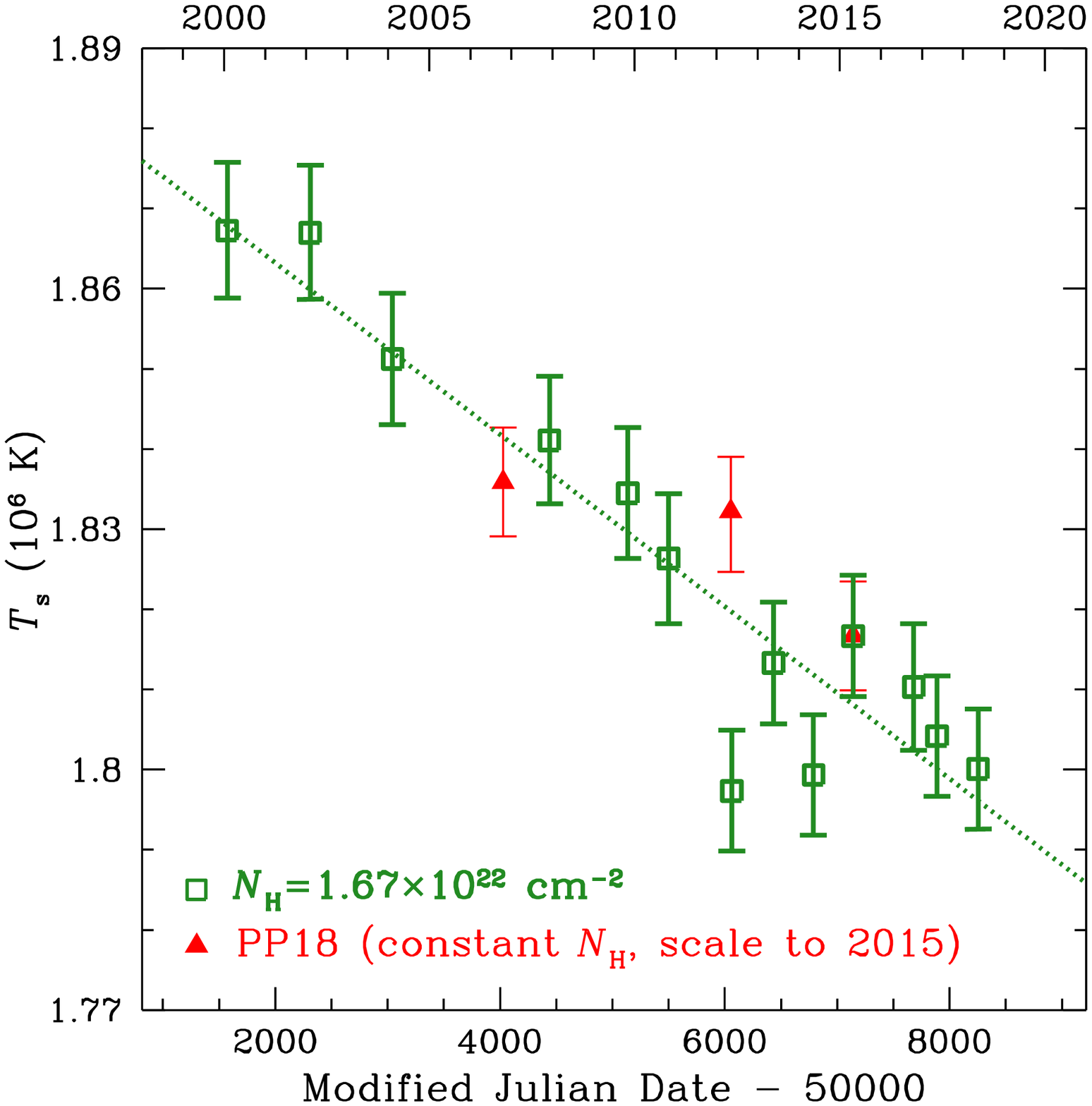}
\includegraphics[width=0.49\textwidth]{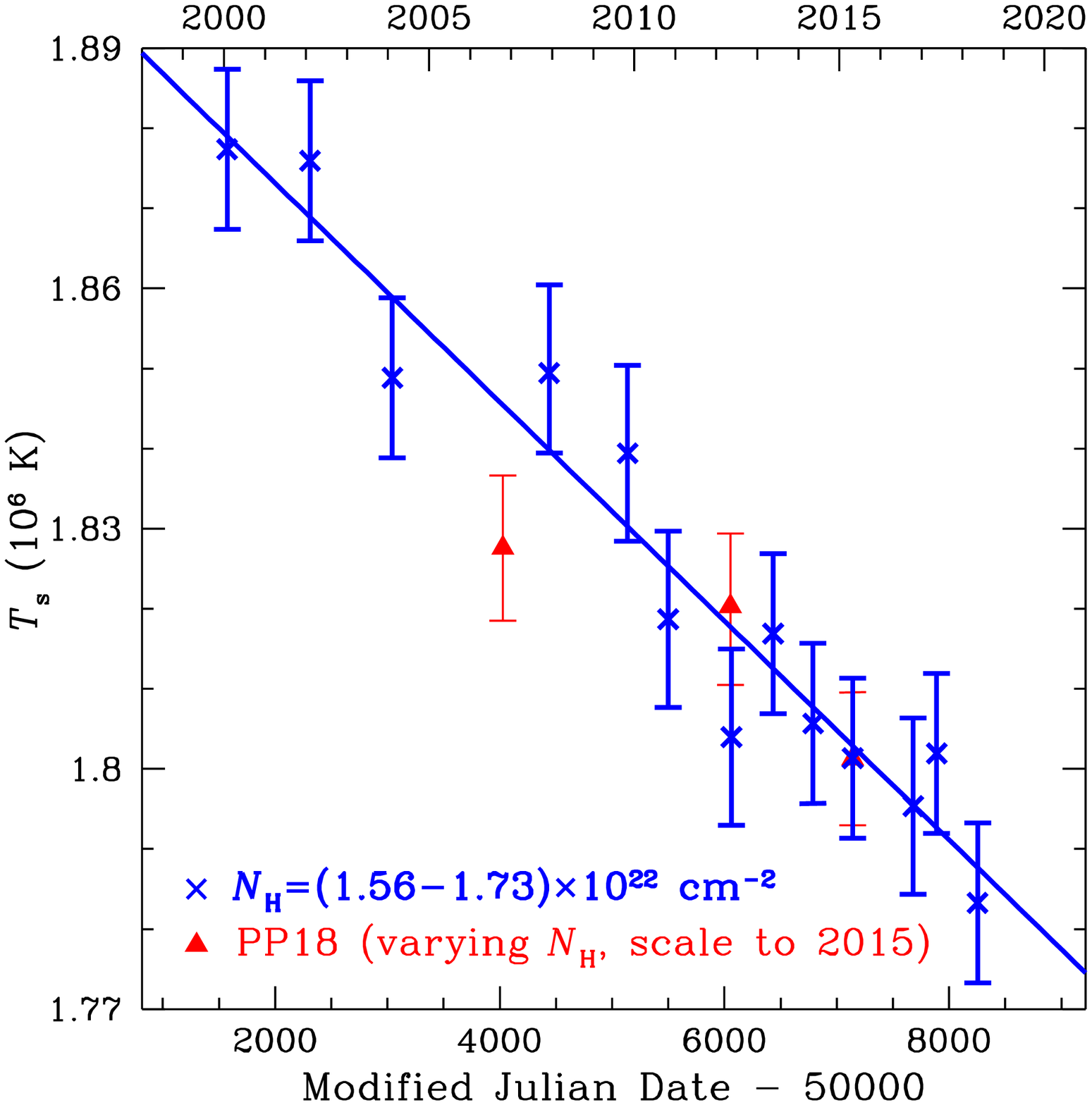}
}
\caption{
Surface temperature of the neutron star in the Cassiopeia~A supernova remnant.
Squares and crosses indicate $\Ts$ measured using \Chandra\ ACIS-S spectra
and best-fit neutron star mass $M=1.65\,\Msun$ and radius $R=12.9\mbox{ km}$.
Error bars are 1$\sigma$.
Lines show a linear fit with a slope of temperature change per decade of
2.1\% (left) and 2.7\% (right).
Left: Interstellar absorption column $\NH$ is held constant when fitting
13 observations over 18 years.
Right: $\NH$ is allowed to vary between each observation.
Triangles are 3 values of $\Ts$ measured using \Chandra\ ACIS-S subarray
spectra by PP18 \cite{posseltpavlov18};
because of differences in fit parameters $M$, $R$, and $\NH$, as well as
uncertainty in instrument cross-calibration, these data are normalized
by the observations in 2015, which were taken less than 3 days apart.
\label{fig:temp}
}
\end{figure}

\begin{figure}[!tb]
\centerline{\includegraphics[width=0.5\textwidth]{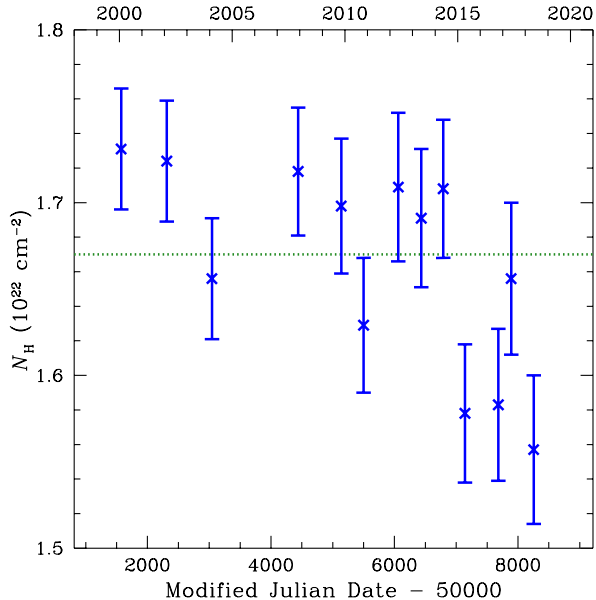}}
\caption{
Interstellar X-ray absorption column $\NH$ in the direction of the
Cassiopeia~A neutron star as measured using \Chandra\ spectra.
Crosses indicate $\NH$ obtained using best-fit neutron star mass
$M=1.65\,\Msun$ and radius $R=12.9\mbox{ km}$ and allowing $\Ts$ to vary,
while the horizontal line indicates the constant value
($\NH=1.67\times 10^{22}\mbox{ cm$^{-2}$}$)
used in the other spectral fit (see text and Figure~\ref{fig:temp}).
Error bars are 1$\sigma$.
\label{fig:nh}
}
\end{figure}

Following its neutron star identification, successive surface temperature
measurements from \Chandra\ spectra show that the neutron star is cooling
at a rapid rate \cite{heinkeho10,elshamoutyetal13,hoetal15,wijngaardenetal19},
likely due to the earlier onset of superconductivity and superfluidity in
the neutron star core \cite{pageetal11,shterninetal11}.
Figure~\ref{fig:temp} shows our latest measurements using
13 \Chandra\ ACIS-S Graded observations over 18 years,
which yield ten-year cooling rates of $2.1\pm0.2\%$ (1$\sigma$) for
constant $\NH$ and $2.7\pm0.3\%$ (1$\sigma$) for varying $\NH$
\cite{wijngaardenetal19}, where $\NH$ measures the column of interstellar
material between us and the neutron star that can absorb X-ray photons.
In performing our spectral fits,
we either fix $\NH$ between each observation to be the same value
[best-fit $\NH=(1.67\pm0.03)\times 10^{22}\mbox{ cm$^{-2}$}$]
or allow $\NH$ to vary between each observation (see Figure~\ref{fig:nh}).
Importantly, our cooling rates agree with the latest findings which use a
\Chandra\ detector configuration (i.e., ACIS-S subarray) that is
optimized for studying the Cassiopeia~A neutron star.
Using 2 of these observations over 6 years, \cite{posseltetal13} find
no cooling ($1.3\pm1.0\%$; 90\% confidence),
and \cite{posseltpavlov18} use a third observation that expands the time
baseline to 9 years to find ten-year cooling rate upper limits of 2.4\%
for constant $\NH$ and 3.3\% for varying $\NH$ (both at 3$\sigma$).
To illustrate the agreement between both sets of observations, we show
temperatures measured by \cite{posseltpavlov18} in Figure~\ref{fig:temp}.
Because of different values of $M$ and $R$ assumed in
\cite{posseltpavlov18}, as well as uncertainty in instrument
cross-calibration, the absolute normalization is somewhat arbitrary.
Therefore, we normalized the temperatures from \cite{posseltpavlov18}
such that temperatures measured in 2015 from observations using the
two different \Chandra\ detector configurations taken less than three
days apart have the same value.

\section{DISCUSSION}

Comparisons between observations of neutron stars and theoretical
modeling of neutron star cooling provide us a unique method to
understand properties of fundamental physics such as the nuclear
EOS and superconductivity and superfluidity.
Our recent work \cite{wijngaardenetal19}, summarized here, examines
the effect DNB has on the outer layers, spectra, and cooling behavior
of isolated neutron stars.
Work studying DNB in accreting neutron star systems is in progress
\cite{wijngaardenetal}.
In addition, we use 13 observations made over 18 years by \Chandra\ of
the (youngest known) neutron star in the Cassiopeia~A supernova remnant
to show that this neutron star is cooling at a rapid rate of 2--3\%
per decade in temperature.
Previous works use this cooling rate to infer the density dependence
of the critical temperatures for proton superconductivity and neutron
superfluidity, which are then compared to theoretically-predicted models
\cite{pageetal11,shterninetal11,hoetal15}.
Future work will use the new results to further improve our knowledge of
nuclear physics properties
(see, e.g., \cite{yakovlevetal11,shterninyakovlev15}).

Meanwhile, more \Chandra\ observations of Cassiopeia~A (and other neutron
stars) will be made.  An important issue that is the subject of ongoing
study is calibration of ACIS detectors onboard \Chandra.
It is known that a contaminant is building up on these detectors,
and this contaminant affects temperature measurements of the
Cassiopeia~A neutron star \cite{elshamoutyetal13,posseltetal13}.
Our recent results \cite{wijngaardenetal19}, as well as those of
\cite{posseltpavlov18}, use the latest models that seek to offset
the effects of the contaminant \cite{plucinskyetal16,plucinskyetal18}.
While the measured variation of $\NH$ with time (see Figure~\ref{fig:nh})
could be due to changes in the amount of intervening X-ray absorbing
material (see, e.g., \cite{alpetal18}), this variation could alternatively
be due to incomplete modeling of the contaminant.
It would be interesting to determine the nature of this variation,
as it affects the cooling rate determination (e.g., 2.1\% for constant
$\NH$ versus 2.7\% for varying $\NH$).
With continued work, future \Chandra\ observations will become more
reliable, and neutron star temperature measurements will be more
accurate.

\section{ACKNOWLEDGMENTS}
WCGH is grateful to the organizers of the Xiamen-CUSTIPEN workshop for
their support and hospitality
and to the referee for their valuable comments.
WCGH thanks the staff at the Chandra X-ray Center for their operation
and maintenance of \Chandra.
WCGH acknowledges support through grant ST/R00045X/1 from Science and
Technology Facilities Council in the United Kingdom.

\nocite{*}
\bibliographystyle{aipnum-cp}
\bibliography{hoetal19_1}

\end{document}